\documentclass[aps,groupedaddress,showpacs,epsfig,graphics]{revtex4}

\usepackage{epsfig}

\begin{document}
\bibliographystyle{apsrev}

\title{Towards a Macroscopic Modelling of the Complexity in Traffic Flow}

\author{Stephan Rosswog$^a$ and Peter Wagner}
\affiliation{German Aerospace Center (DLR), 51170 K\"oln-Porz, Germany}
\affiliation{$^a$ On leave: }

\date{\today}
\begin{abstract}
  Based on the assumption of a safe velocity $U_{\rm
  e}(\rho)$ depending on the vehicle density $\rho$ 
  a macroscopic model for traffic flow is presented that
  extends the model of the K\"uhne-Kerner-Konh\"auser 
  by an interaction term containing the second derivative
  of $U_{\rm e}(\rho)$. We explore two qualitatively different 
  forms of $U_{\rm e}$: a conventional, Fermi-type function
  and, motivated by recent experimental findings, a function
  that exhibits a plateau at intermediate densities, i.e.\ 
  in this density regime the exact
  distance to the car ahead is only of minor importance.
  To solve the fluid-like equations a Lagrangian particle scheme
  is developed.\\
  The suggested model shows a much richer dynamical behaviour than the
  usual fluid-like models. A large variety of encountered effects is
  known from traffic observations many of which are usually assigned
  to the elusive state of ``synchronized flow''. Furthermore, the model
  displays alternating regimes of stability and instability at intermediate
  densities, it can explain data scatter in the fundamental diagram
  and complicated jam patterns.
  Within this model, a consistent interpretation of the emergence of 
  very different traffic phenomena is offered: they are determined by 
  the velocity relaxation time, i.e. the time needed to relax towards 
  $U_{\rm e}(\rho)$.
  This relaxation time is a measure of the average acceleration capability
  and can be attributed to the composition (e.g. the percentage of trucks)
  of the traffic flow. 
\end{abstract}
\maketitle

\section{Introduction}

Traffic is a realization of an open one-dimensional many-body
system. Recently, Popkov and Sch\"utz \cite{popkov99} found that the
fundamental diagram determines the phase diagram of such a system, at
least for a very simple, yet exactly solvable toy model, the so called
asymmetric exclusion process (ASEP). In particular, the most important
feature that influences the phase diagram is the number of extrema in
the fundamental diagram.
  
This is exactly the theme of this report. We present an extension of
classical, macroscopic (``fluid-like'') traffic flow models.  Usually,
it is assumed that the fundamental diagram is a one-hump function,
however recent empirical results point to  more complicated
behaviour. It is impossible to assign a single flow function $j(\rho)$
to the measured data-points in a certain density range. Therefore, it
can be speculated, that this scatter hides a more complicated
behaviour of the fundamental diagram in this regime.  
We explore two qualitatively different forms of the safe velocity $U_e(\rho)$,
the velocity to which the flow tends to relax, which leads from the usual
one-hump behaviour of the flow density relation to a more complicated 
function that exhibits, depending on the relaxation parameter, one, two or 
three humps.
Obviously, real drivers may have different $U_e(\rho)$--functions,
adding another source of dynamical complexity, which will not be
discussed in this paper.
 
\section{The Model}
\subsection{Equations}
If the behaviour of individual vehicles is not of concern, but the
focus is more on aggregated quantities (like density $\rho$, mean
velocity $v$ etc.), one often describes the system dynamics by means of
macroscopic, fluid-like equations. The form of these
Navier-Stokes-like equations can be motivated from 
anticipative behaviour of the drivers.\\
Assume there is a safe velocity $U_e$ that only depends on the density $\rho$.
The driver is expected to adapt the velocity in a way that $v$ relaxes on 
a time scale $\tau$ to this desired velocity corresponding to the density at
$x + \Delta x$,
\begin{equation}
v(x+v \tau,t+\tau)= U_e(\rho(x+ \Delta x))\label{ansatz}.
\end{equation}
If both sides are Taylor-expanded to first order one finds
\begin{equation}
v(x)+ \frac{\partial v}{\partial x} v \tau + \frac{\partial v}
{\partial t} \tau + O(\tau^2) = U_e(\rho) + \frac{\partial U_e}{\partial  \rho}
\frac{\partial \rho}{\partial x} \Delta x + O((\Delta x)^2). 
\end{equation}
%
Inserting $\Delta x = \rho^{-1}$
\begin{equation}
 \frac{\partial v}{\partial t} + v \frac{\partial v}{\partial x}
= \frac{U_e(\rho)-v}{\tau}+\frac{1}{\rho \tau} \frac{\partial 
U_e(\rho)} {\partial \rho} \frac{\partial \rho}{\partial x}.
\end{equation}
Abbreviating $\frac{\partial U_e(\rho)} {\partial \rho} 
\frac{1}{\tau}$ with $-c_0^2$ the Payne 
equation \cite{payne71} is recovered: 
\begin{equation}
 \frac{\partial v}{\partial t} + v \frac{\partial v}{\partial x}
= \frac{U_e(\rho)-v}{\tau} - \frac{c_0^2}{\rho} 
\frac{\partial \rho}{\partial x}.
\end{equation}
If one seeks the analogy to the hydrodynamic equations one can
identify a ``traffic pressure'' $P= c_0^2 \rho$. In this sense traffic
follows the equation of state of a perfect gas (compare to
thermodynamics: $P= n k_B T$).\\
The above described procedure to
motivate fluid-like models can be extended beyond the described model
in a straight forward way.  If, for example, eq. (\ref{ansatz}) is
expanded to second order, quadratic terms in $\tau$ are neglected, the
abbreviation $c_0$ is used and the terms in front of $\frac{\partial^2
U_e}{\partial \rho^2}$ are absorbed in the coupling constant $g$, one
finds:
\begin{equation}
 \frac{\partial v}{\partial t} + v \frac{\partial v}{\partial x}
= \frac{U_e(\rho)-v}{\tau} - \frac{c_0^2}{\rho}
\frac{\partial \rho}{\partial x} + g\; U''_e(\rho).
\end{equation}
The primes in the last equation denote derivatives with respect to the
density. Since these equations allow infinitely steep velocity
changes, we add (as in the usual macroscopic traffic flow equations
\cite{kuehne84},\cite{kerner93}) a diffusive term to smooth out shock
fronts:
\begin{equation}
 \frac{\partial v}{\partial t} + v \frac{\partial v}{\partial x}
= \frac{U_e(\rho)-v}{\tau} - \frac{c_0^2}{\rho}
\frac{\partial \rho}{\partial x} 
+ \frac{\mu}{\rho} \frac{\partial^2 v}{\partial x^2} + g\;  U''_e(\rho) 
\label{modeq} .
\end{equation}
Since a vehicle passing through an infinitely steep velocity shock
front would suffer an infinite acceleration, we interpret the
diffusive (``viscosity'') term as a result of the finite acceleration
capabilities of real world vehicles.  Our model equations
(\ref{modeq}) extend the equations of the K\"uhne-Kerner-Konh\"auser
(in the sequel called K$^3$ model; \cite{kuehne84},\cite{kerner93}) model by a term
coupling to the second derivative of the desired velocity. Throughout
this study we use $c_0= 15$ ms$^{-1}$, $\mu= 50$ ms$^{-1}$ and $g= 8
\cdot 10^{-4}$ m$^{-2}$s$^{-1}$.
\subsection{Shape of the safe velocity}
\begin{figure}[h]
\begin{center}
\epsfig{file=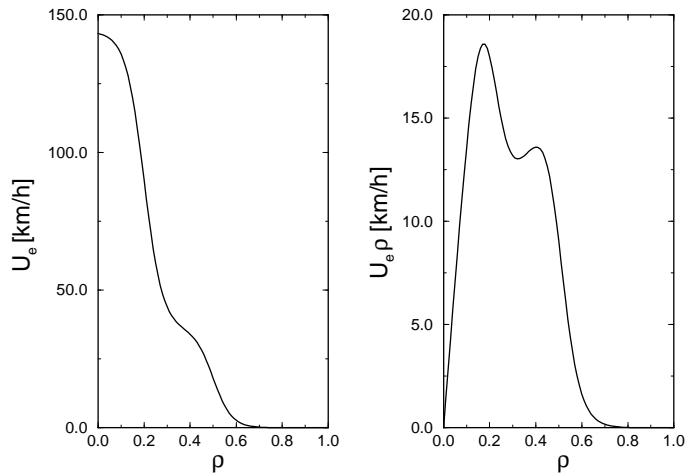,angle=-90,width=0.5\linewidth}
\caption{\label{Vdes} Safe velocity with a plateau and the corresponding flow.
For details see text.}
\end{center}
\end{figure}

The form of the safe velocity $U_e$ plays an important role
in this class of models (as can be seen, for example, from the linear
stability analysis of the $K^3$ model). However, experimentally the
relation between this desired velocity and the vehicle density is
poorly known. It is reasonable to assume a maximum  at vanishing
density and once the vehicle bumpers touch, the velocity will
(hopefully) be zero.\\
To study the effect of the additional term in the equations of motion 
we first investigate the case of the conventional safe velocity given
by a Fermi-function of the form \cite{kerner93}
\begin{equation}
U_e= 40\left(1/(1+exp((\rho-0.25)/0.06))-3.72\cdot 10^{-6}\right)
\mbox{ms}^{-1}.
\end{equation}
Since $U_e$ is at present stage rather
uncertain, we also examine  the effects of a more complicated
relation between the desired velocity $U_e$ and the density
$\rho$. For this reason we look at a velocity-density relation that
has a plateau at intermediate densities, which, in a microscopic
interpretation, means that in a certain density regime drivers do not
care about the exact distance to the car ahead.  We chose an
$U_e$-function of the form
\begin{equation}
U_e(\rho) = \frac{V_0}{2} (\xi(\rho) - \xi(1)), \qquad \rho \in [0,1] 
\label{Vdes_eq}
\end{equation}
with
\begin{equation}
\xi(x) = n_1 \, \Theta(x - x_{\rm c1}) + n_2 \, \Theta(x - x_{\rm c2})
\end{equation}
where $\Theta(x) = 1/(1+exp(a\,x))$ is used.  The parameters $n_1=1.5,
n_2=0.5, a=25.0, x_{\rm c1}=0.2$, $x_{\rm c1}=0.5$ and $V_0= 40$ m
s$^{-1}$ are used throughout this study, the corresponding safe velocity and 
flow are shown in Fig.~\ref{Vdes}. Note that the densities are
always normalized with respect to their maximum possible value
$\rho_{max}$ which is given by the average vehicle length as
$l_{veh}^{-1}$.\\

\section{The numerical method: a Lagrangian particle scheme}
We use a Lagrangian particle scheme
to solve the Navier-Stokes-like equations for traffic flow. 
A particle method similar to the smoothed 
particle hydrodynamics method (SPH; \cite{Benz90}) has been used 
previously to simulate traffic flow 
\cite{rosswog99}, the method we use here, however,
differs in the way the density and the derivatives are calculated.
The particles correspond to moving
interpolation centers that carry aggregated properties of the vehicle
flow, like, for example, the vehicle density $\rho$.  They are not to
be confused with single ``test vehicles'' in the flow, they rather
correspond to ``a bulk'' of vehicles.\\ 
The first step in this procedure
is to define, what is  meant by the term ``vehicle density''.  Since
we assign a number indicating the corresponding vehicle number $n_i$
to each particle $i$ with position $x_i$, the density definition is
straight forward, i.e. the number of vehicles per length that can be
assigned unambiguously to particle $i$, or

\begin{equation}
\rho_i= \frac{n_i}{(x_{i+1}-x_{i})/2 + (x_{i}-x_{i-1})/2}= 
\frac{2 n_i}{x_{i+1} - x_{i-1}}.\label{dens}
\end{equation}

\begin{figure}[ht]
\begin{center}
\epsfig{file=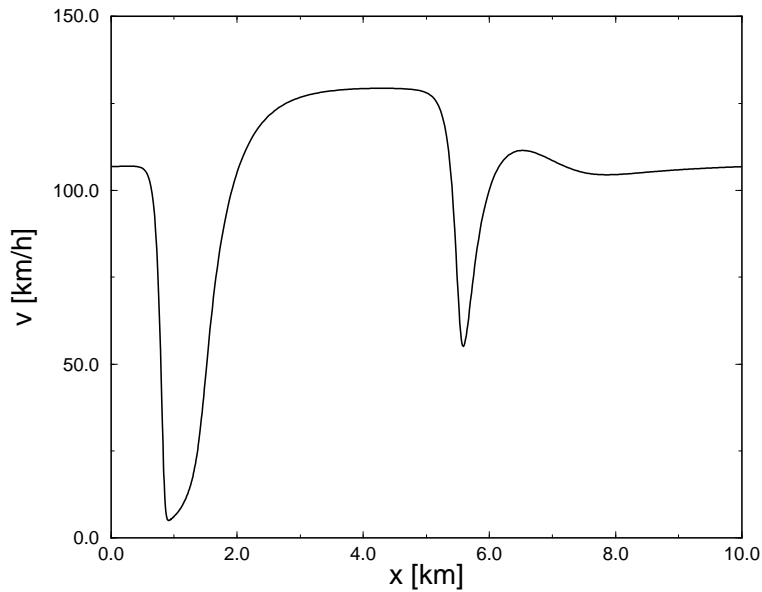,angle=-90,width=10cm}
\caption{\label{shockfront} Emerging spontaneous breakdown of traffic flow
in the unstable regime of the K$^3$-model ($\rho_{in}=0.20$, $\mu=50$
ms$^{-1}$, $\tau=10$ s). Shown is a developed, but still broadening,
backward moving jam and a sharply localized, forward moving and still
steepening velocity perturbation.}
\end{center}
\end{figure}

Once this is done one has to decide in which way spatial derivatives
are to be evaluated. One possibility would be to take finite
differences of properties at the particle positions. However, one has
to keep in mind that the particles are not necessarily distributed
equidistantly and thus in standard finite differences higher order terms
do not automatically cancel out {\em exactly}. The introduced
errors may be appreciable in the surrounding of a shock and they can
trigger numerical instabilities that prevent further integration
of the system.  Therefore we decided to evaluate first order
derivatives as the {\em analytical derivatives} of cubic spline
interpolations through the particle positions. Second order
derivatives of a variable $f$ are evaluated using centered finite
differences

\begin{equation}
\frac{\partial^2 f} {\partial x^2}(x)=\frac{f_{+} + f_{-} - 2 f(x)}{\delta^2} +
O(\delta^3),
\end{equation} 

where $f_{+}\equiv f(x+\delta)$ and $f_{-}\equiv f(x-\delta)$ are
evaluated by spline interpolation and $\delta$ is an appropriately
chosen discretisation length. Since we do not evolve the ``weights''
$n_i$ in time, there is no need to handle a continuity equation, the
total vehicle number $N$ is constant and given as $N= \sum_i n_i.$\\
Denoting the left hand side of (\ref{modeq}) in Lagrangian form
$\dot{v}\equiv \frac{dv}{dt}= \frac{\partial v}{\partial t} + v
\frac{\partial v}{\partial x}$, we are left with a first order system:
\begin{eqnarray}
\dot{x}_i &=& v_i \label{eq1}\\
\dot{v}_i &=& \frac{U_e(\rho)-v}{\tau} - \frac{c_0^2}{\rho}
\frac{\partial \rho}{\partial x} 
+ \frac{\mu}{\rho} \frac{\partial^2 v}{\partial x^2} + g\;  U''_e(\rho).
\label{eq2} 
\end{eqnarray}
This set of equations is integrated forward in time by means of a
fourth order accurate Runge-Kutta integrator with adaptive time
step.\\
The described scheme is able to resolve emerging shock fronts
sharply without any spurious oscillations. An example of such a shock
front is shown in Fig.~\ref{shockfront} for the $K^3$-model.

\section{Complexity in Traffic Flow}

\begin{figure}[hb]
\begin{center}
\epsfig{file=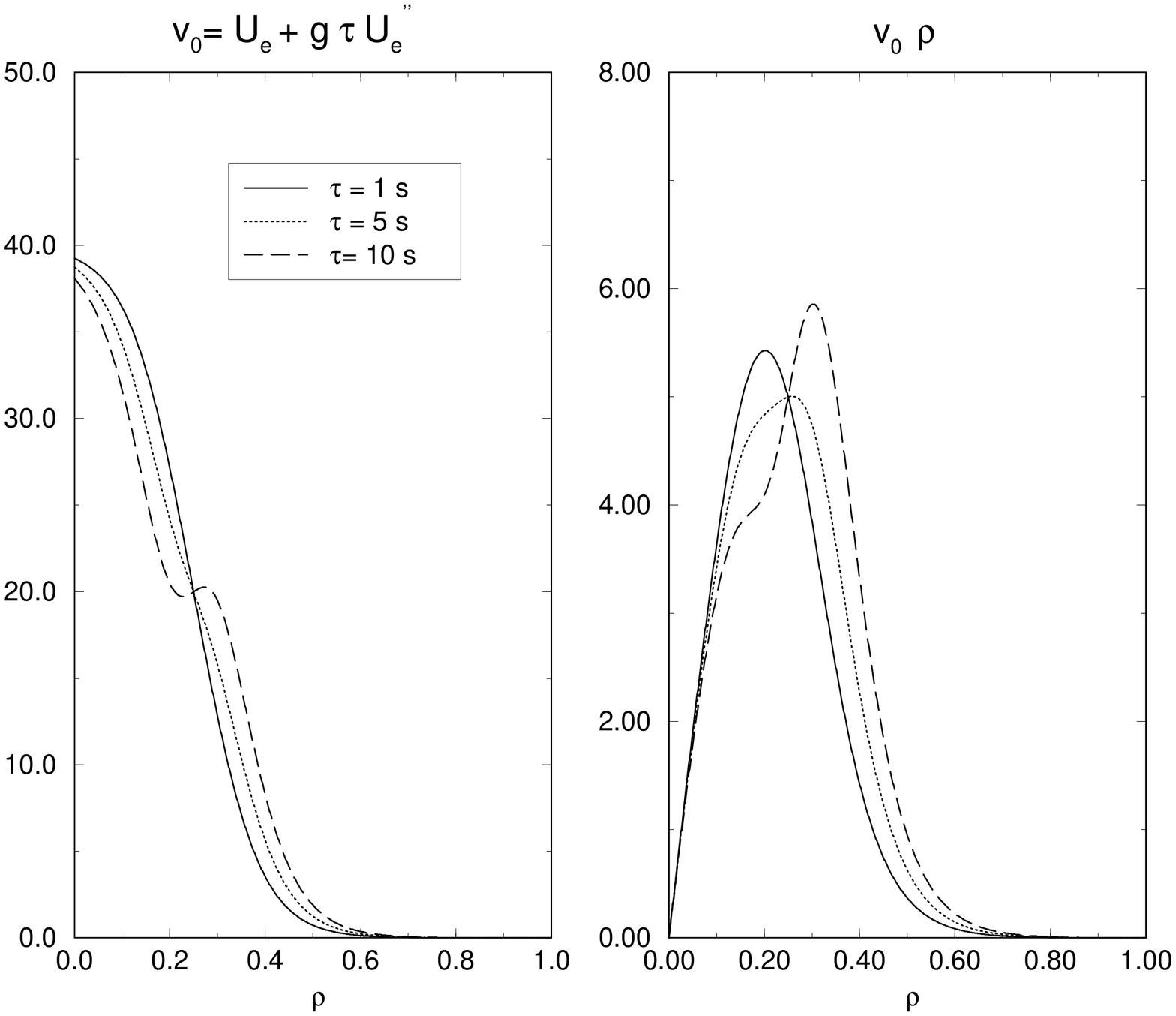,angle=-90,width=8cm}
\epsfig{file=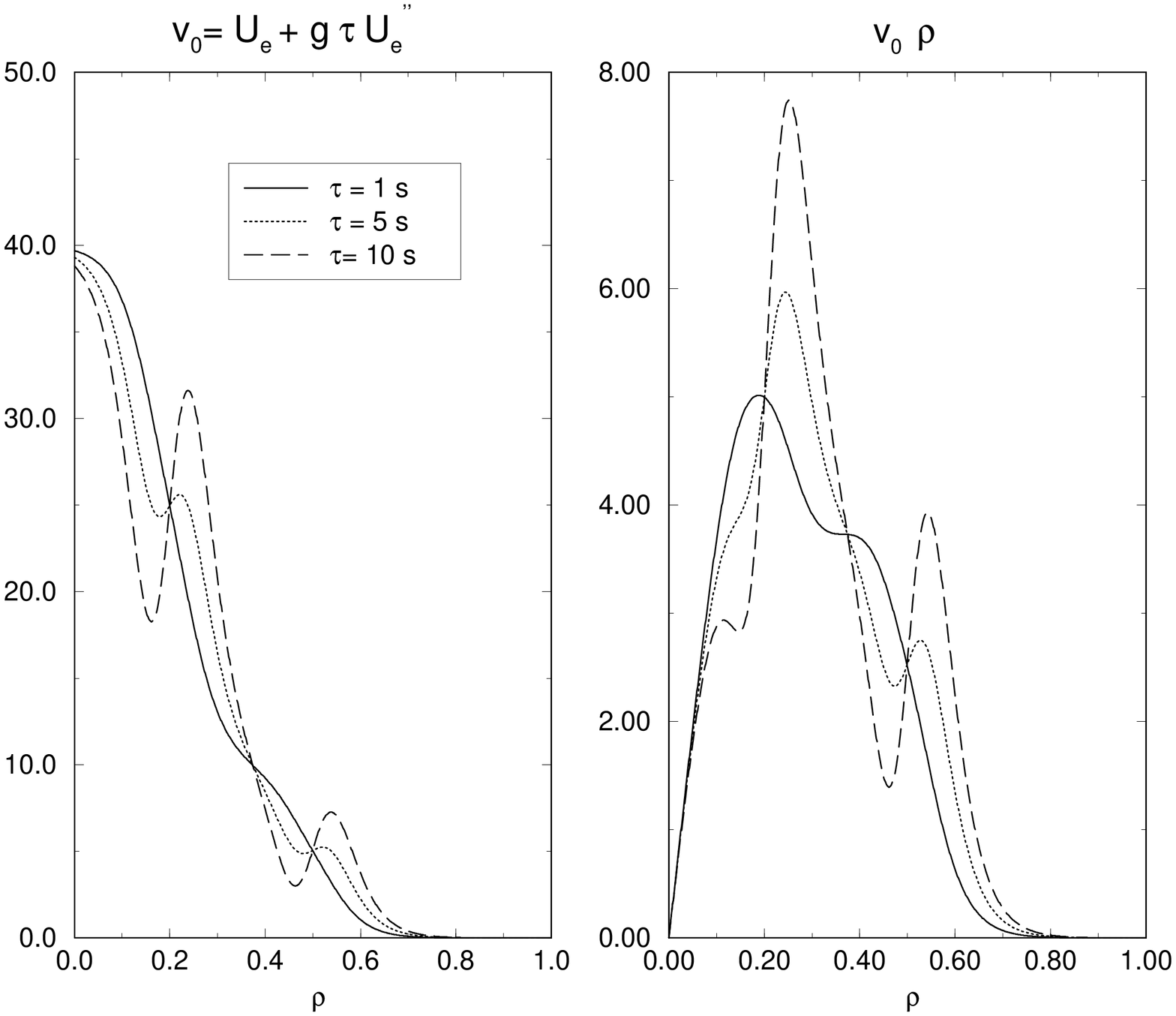,angle=-90,width=8cm}
\caption{ Left two panels: velocity isoclines 
(i.e points where accelerations vanish) for the homogeneous and stationary
solution as a function of the local density $\rho$ for our model equations
and the $U_e$ of the K$^3$ model and the corresponding fundamental diagrams. 
All velocities are measured in ms$^{-1}$.
Right two panels: dito but for the $U_e$ with a plateau.}
\label{force_free_FD}
\end{center}
\end{figure}

\begin{figure*}
\begin{minipage}[]{5cm}
\hspace*{-5cm}\psfig{file=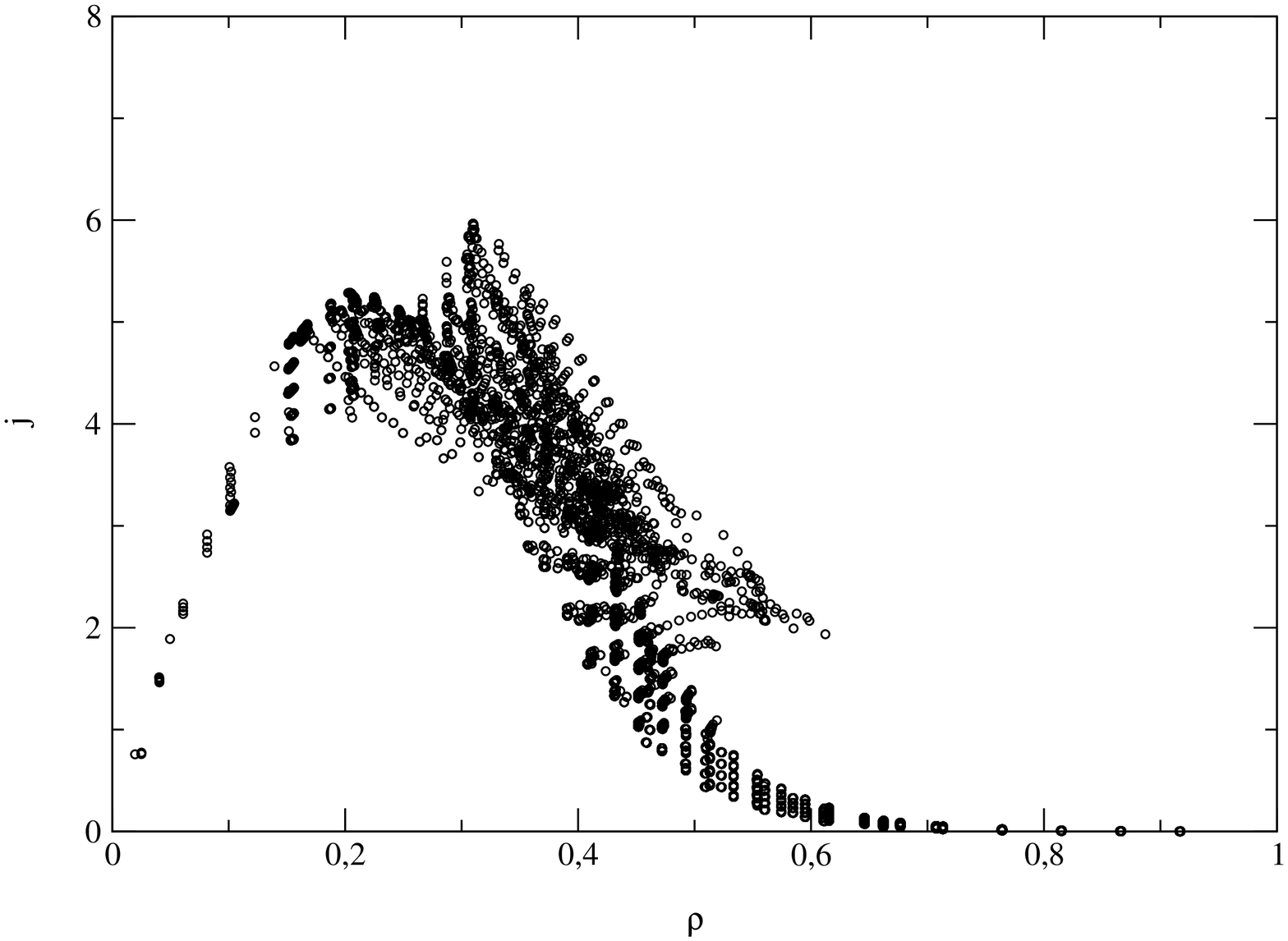,angle=-90,width=10cm}
\end{minipage}
\begin{minipage}[]{5cm}
\psfig{file=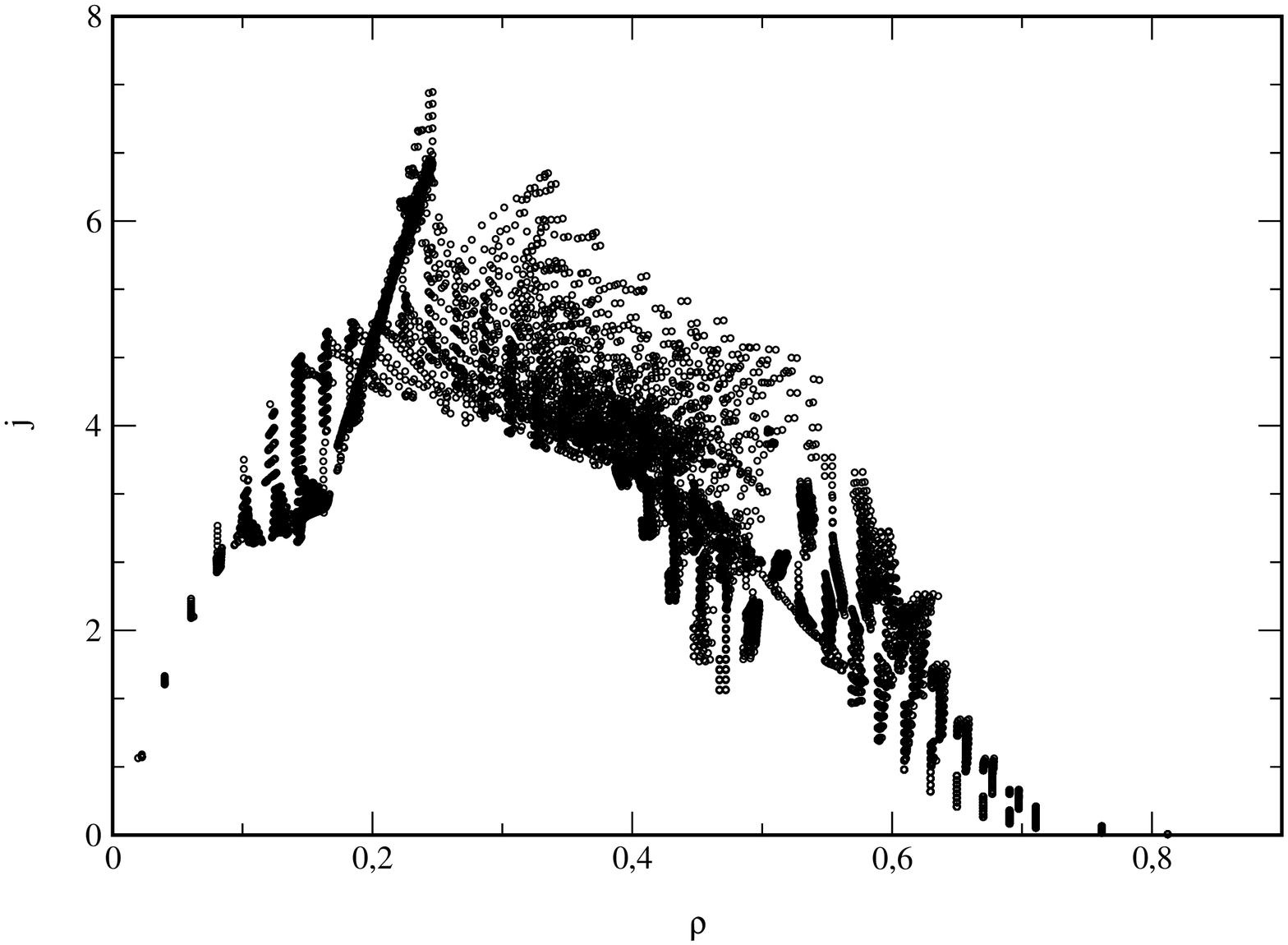,angle=-90,width=10cm}
\end{minipage}
\caption{\label{FD_allTau} Shown are measured 1 minute averages of a set of
simulations where the time scale $\tau$ has been scanned from 1 up to 10 s.
The left panel results from using the conventional form of the safe velocity, 
for the right one the plateau function was used.}
\end{figure*}

\begin{figure}[h]
\begin{center}
\epsfig{file=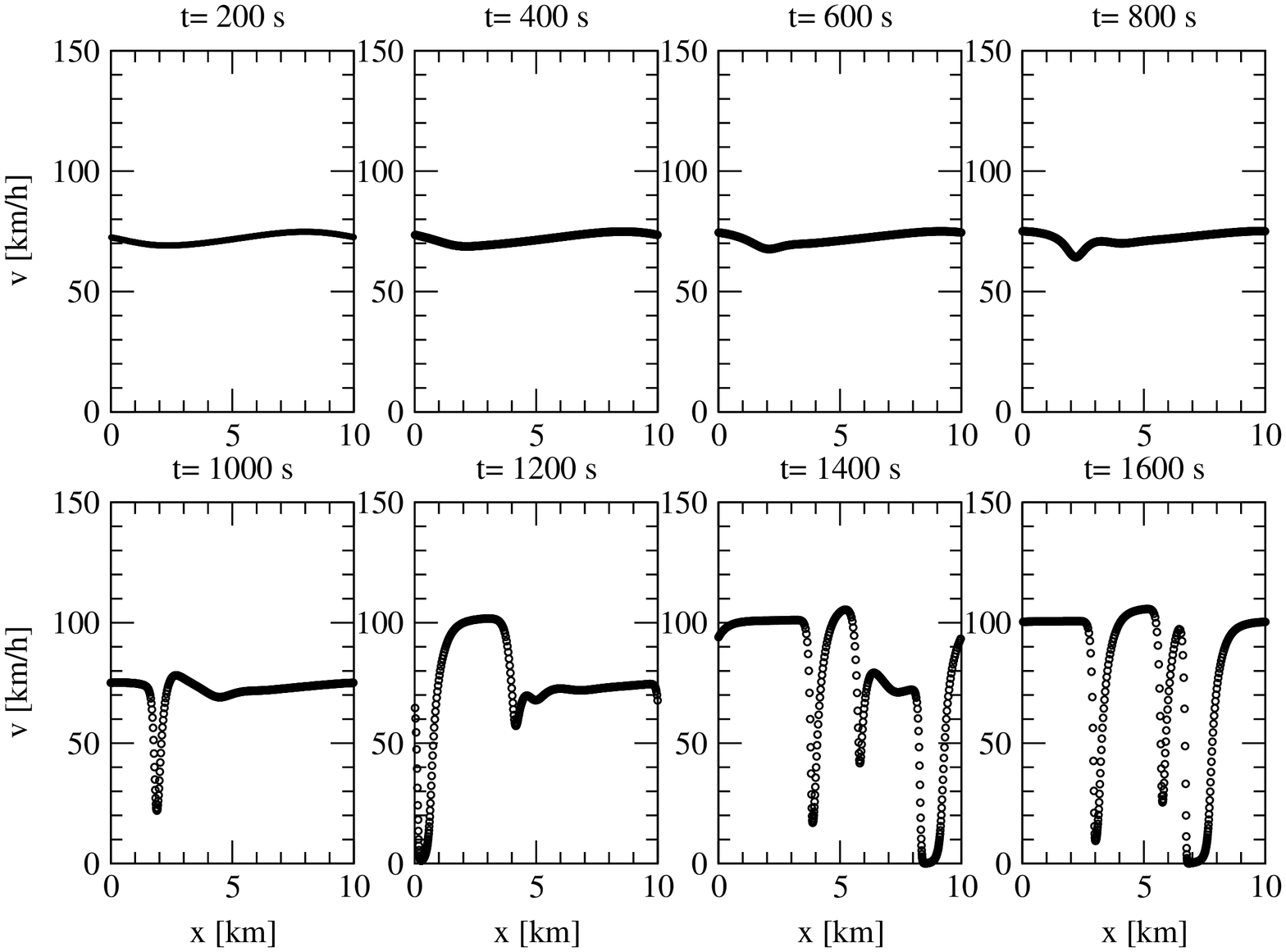,angle=-90,width=8.9cm}
\epsfig{file=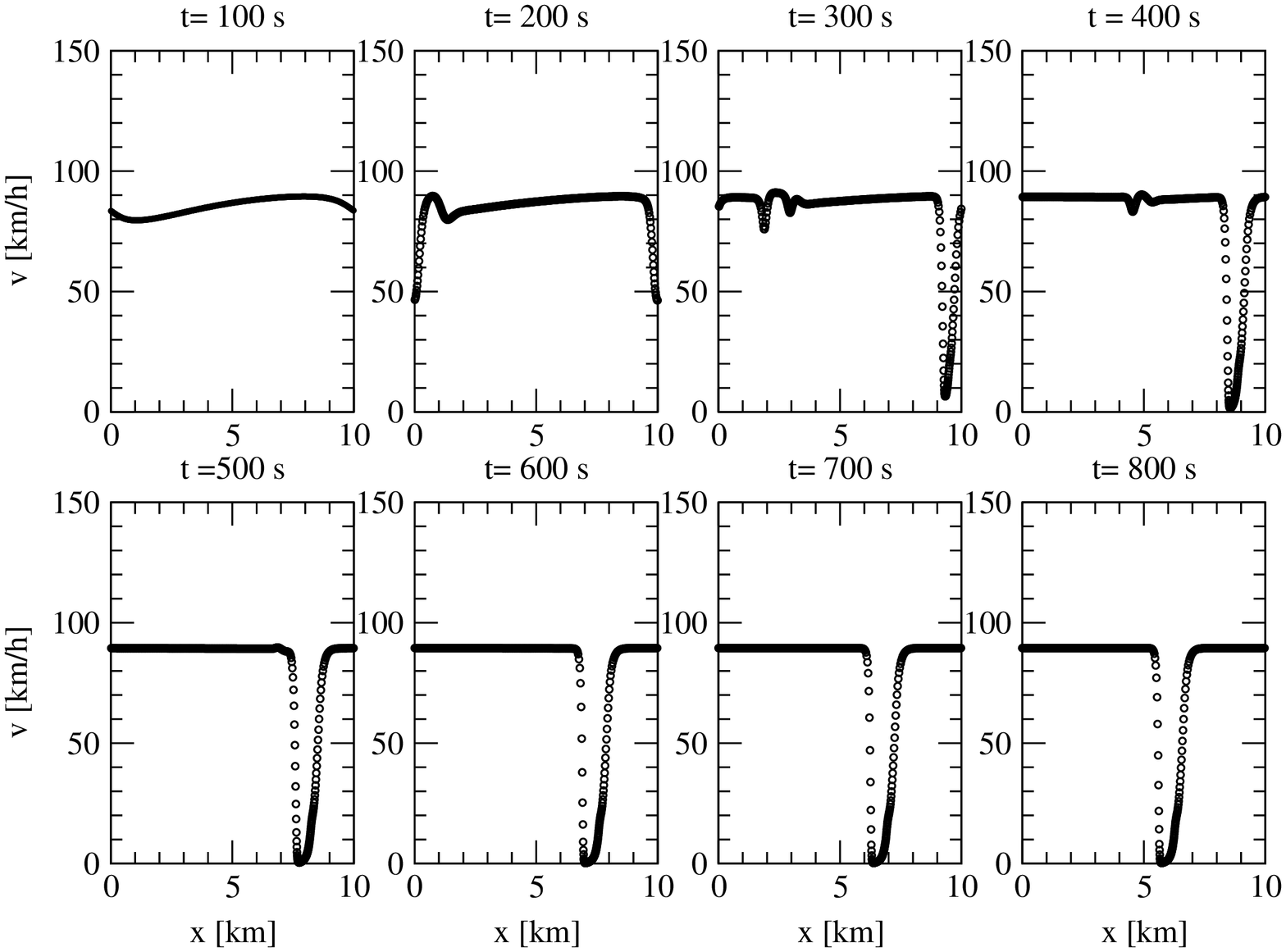,angle=-90,width=8.9cm}
\caption{\label{widejam} Emerging spontaneous breakdown of traffic flow
in the unstable regime of our model ($\mu=50$ ms$^{-1}$ , $\tau=5$ s, 
$\rho_{in}= 0.25$). The growing jam moves with $\approx 28$ km h$^{-1}$ 
backwards.}
\end{center}
\end{figure}

\begin{figure}[t]
\begin{minipage}[]{5cm}
\hspace*{-5cm}\psfig{file=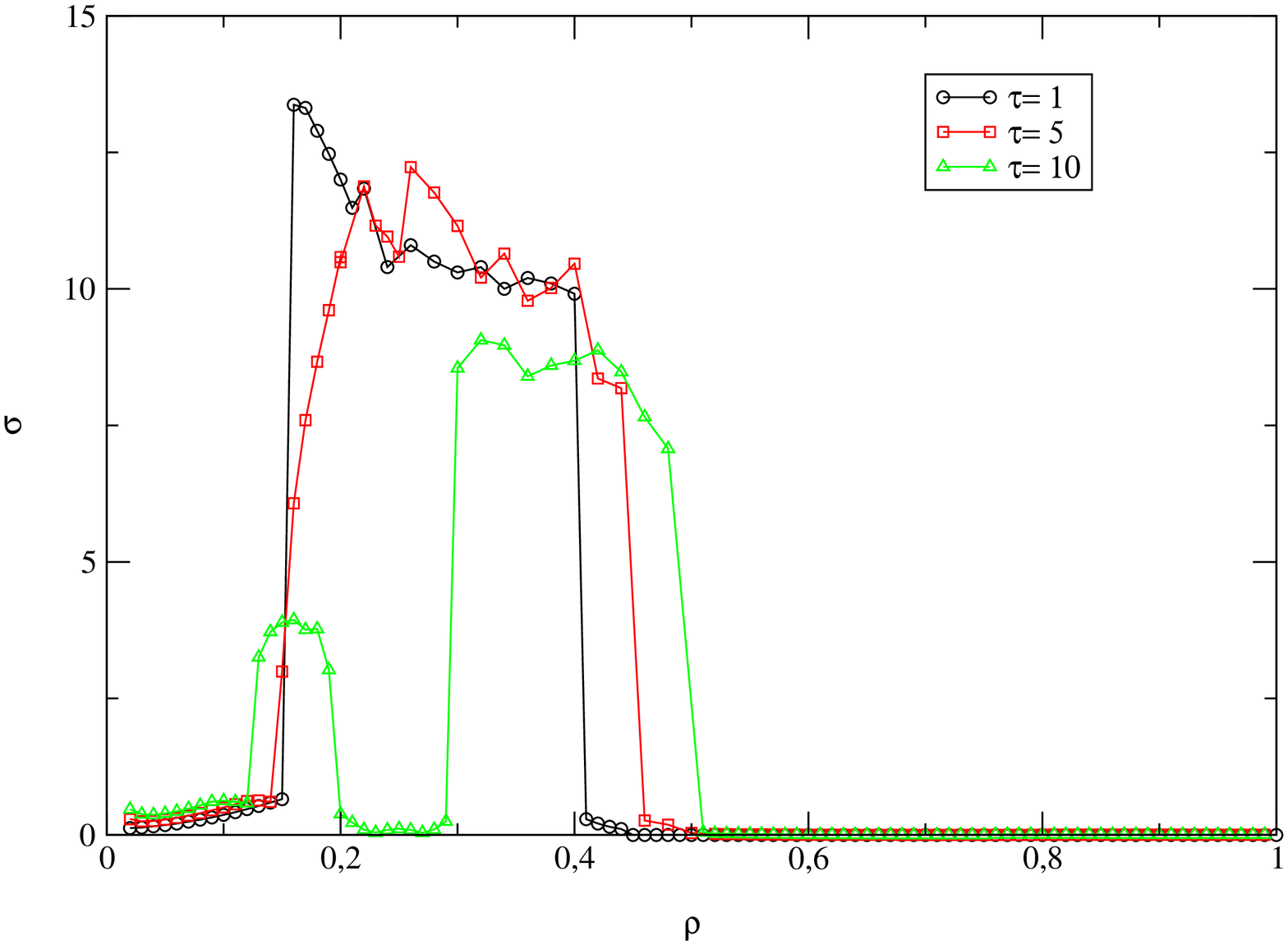,angle=-90,width=9cm}
\end{minipage}
\begin{minipage}[]{5cm}
\vspace*{.3cm}
\epsfig{file=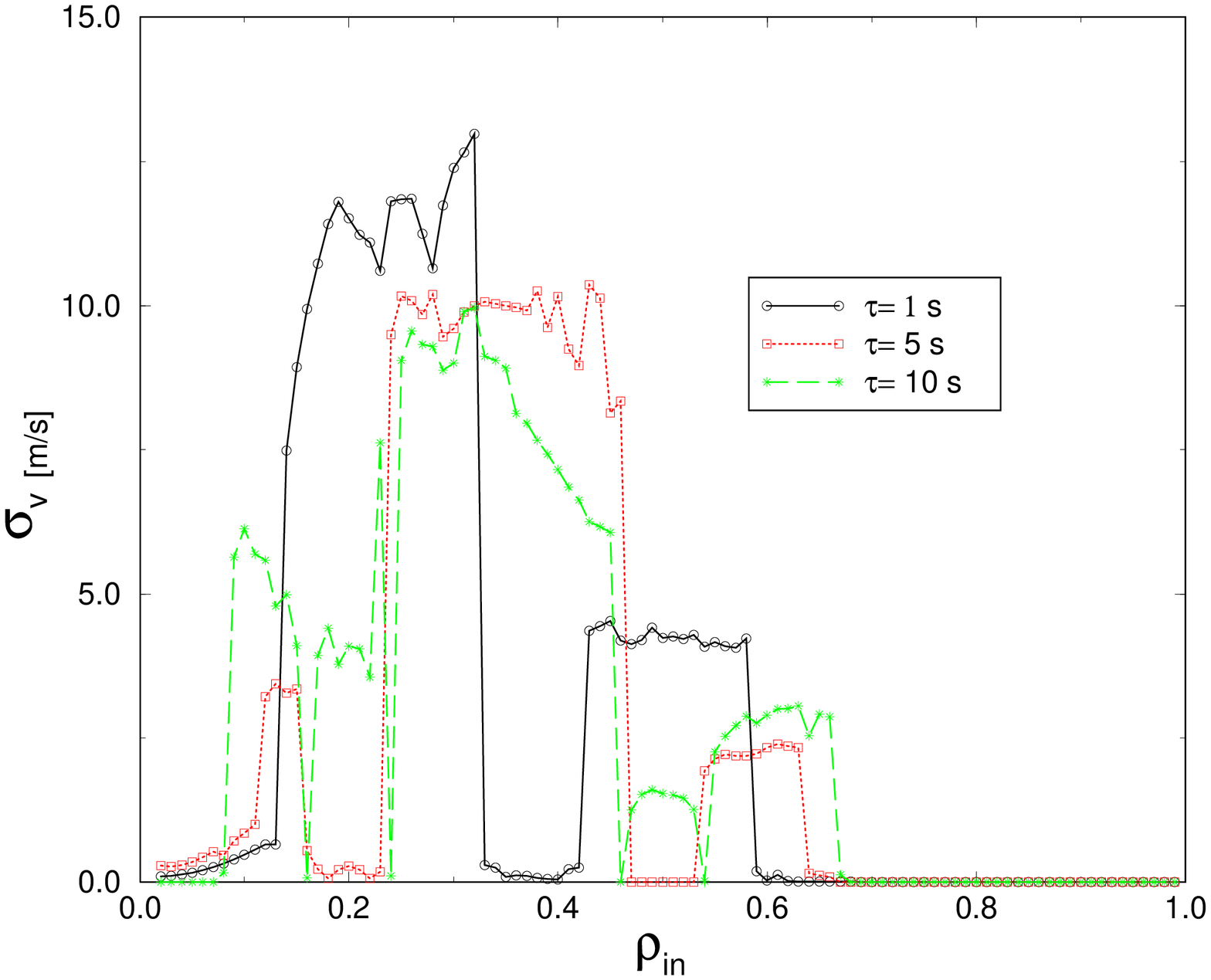,angle=-90,width=7cm}
\end{minipage}
\caption{\label{velvar} Velocity variance $\sigma_v$ 
for $\tau=1, 5, 10$ s as a function of the initial density for the 
conventional form of the safe velocity (left) and the plateau function 
(right).}
\end{figure}

\begin{figure}[h]
\begin{center}
\psfig{file=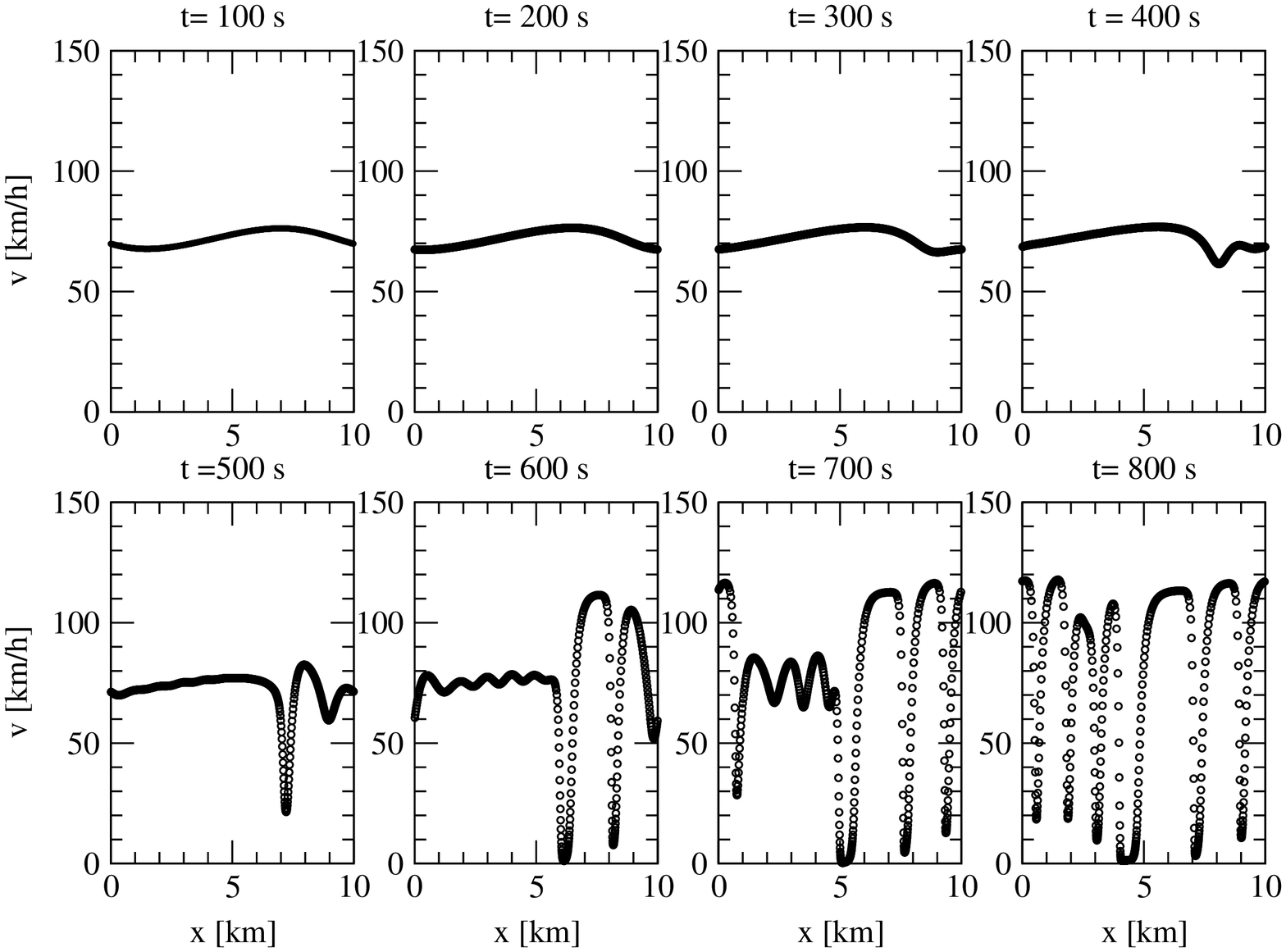,angle=-90,width=8.9cm}
\psfig{file=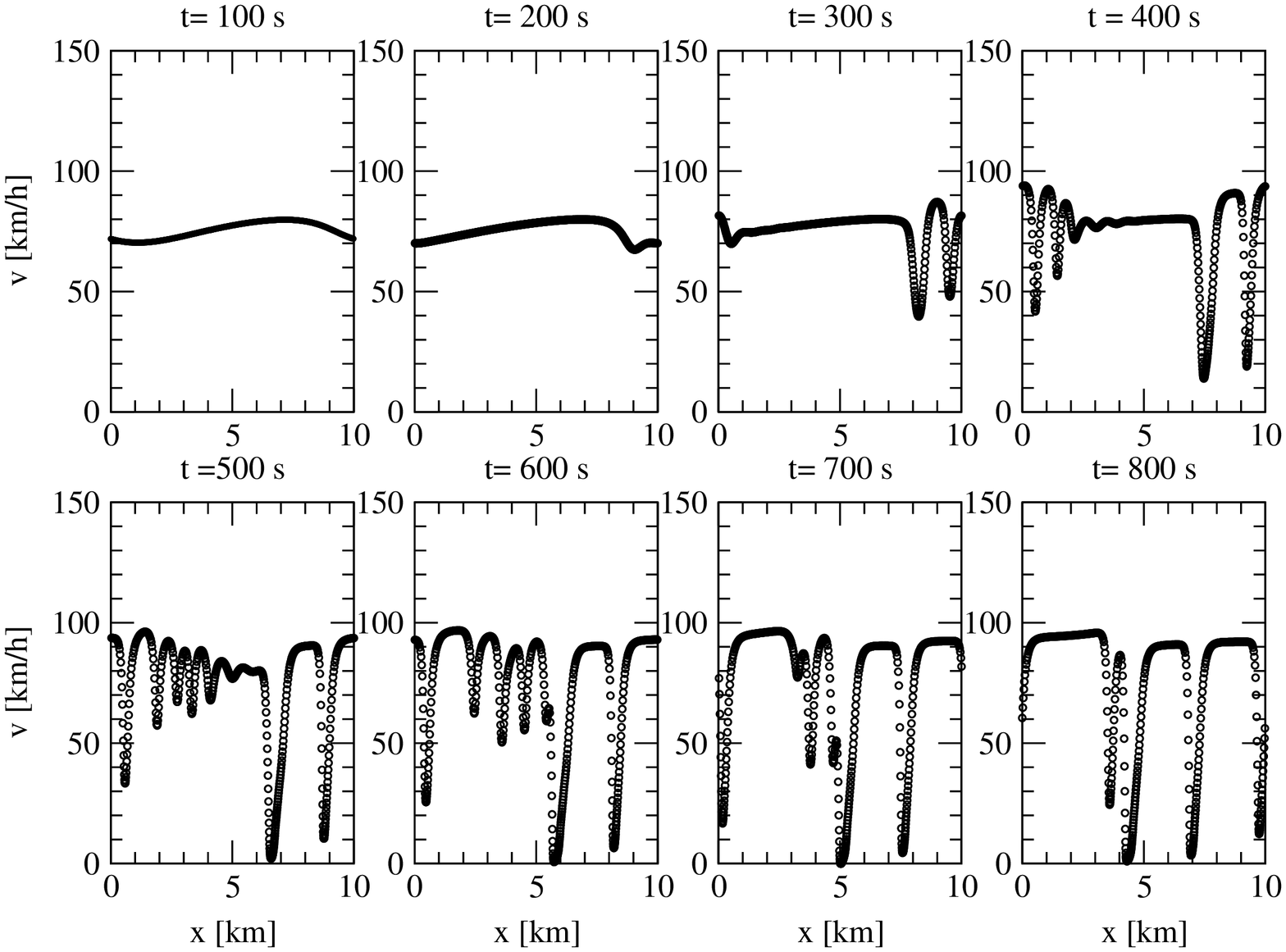,angle=-90,width=8.9cm}
\caption{\label{stop_and_go} Formation of stop-and-go-waves out of nearly
homogeneous initial conditions ($\rho_{in}= 0.25, \mu= 50$ ms$^{-1},
\tau= 3$ s). For the left panel the conventional form of the safe velocity 
was used, the right panel corresponds to the plateau safe velocity. }
\end{center}
\end{figure}

\begin{figure}[b]
\begin{center}
\psfig{file=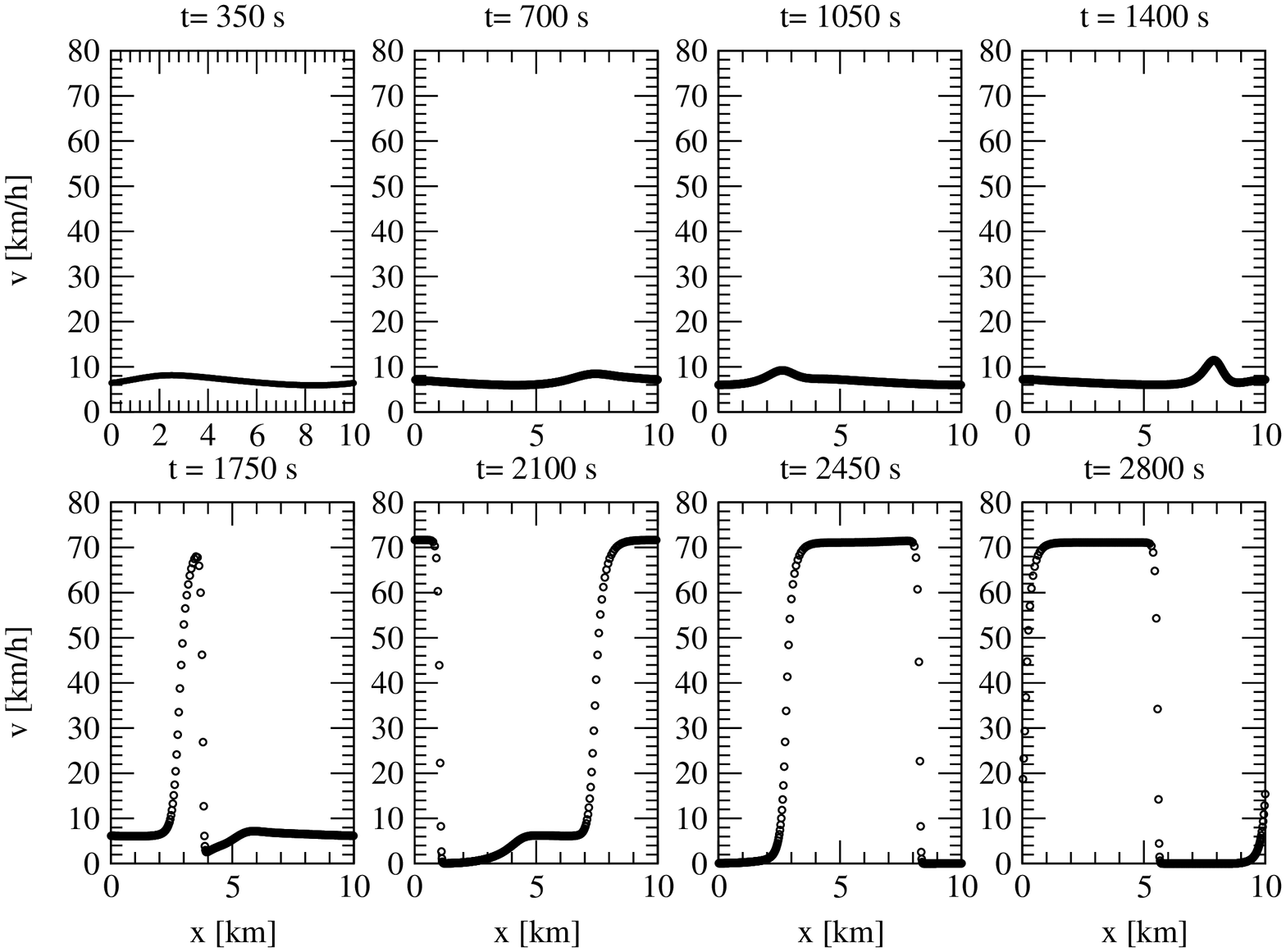,angle=-90,width=8.9cm}
\psfig{file=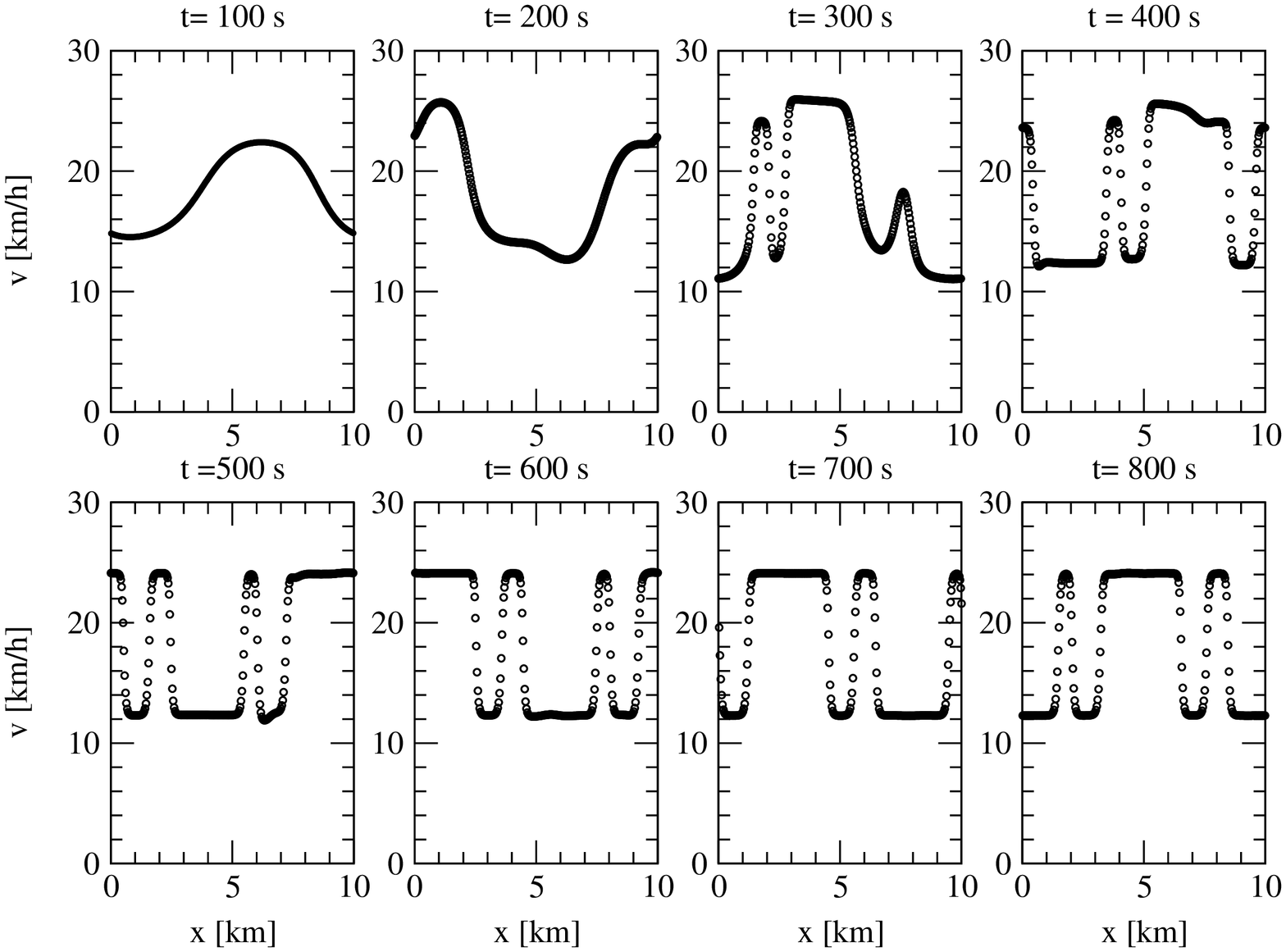,angle=-90,width=8.9cm}
\caption{\label{plateau1} ``Mesa-effect 1'': formation of velocity plateaus 
($\rho_{in}= 0.50, \mu= 50$ ms$^{-1}, \tau= 10$ s) for the conventional (left)
and the plateau safe velocity (right).}
\end{center}
\end{figure}

\begin{figure}[p]
\begin{center}
\epsfig{file=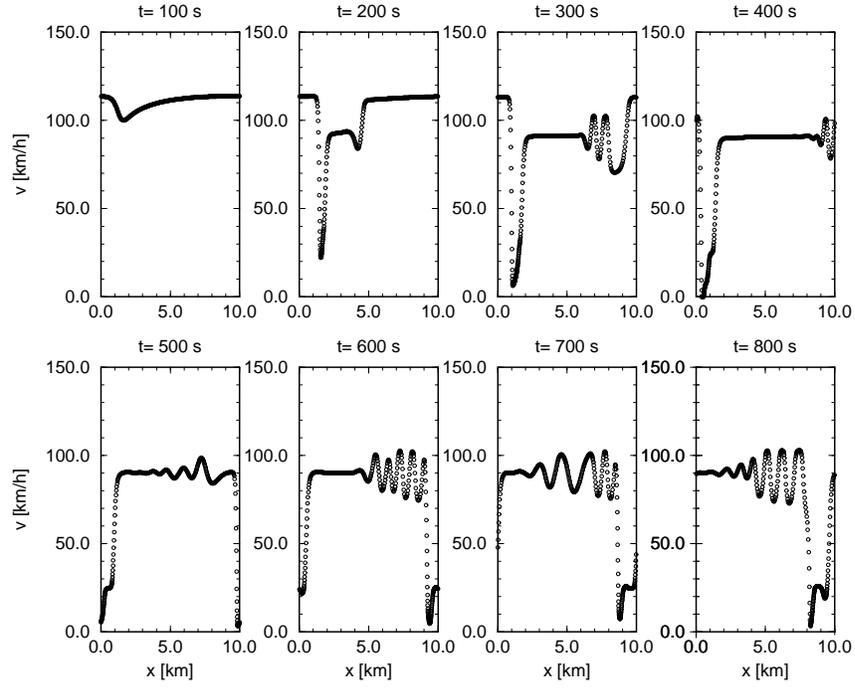,angle=-90,width=11cm}
\caption{\label{jamsync} Emerging spontaneous breakdown of traffic flow
in the unstable regime of our model ($\rho_{in}=0.26$, $\mu=50$ ms$^{-1}, 
\tau=10$ s) with 
the plateau safe velocity. The use of the conventional safe velocity 
(not shown) results in a quick relaxation towards the homogeneous state.}
\end{center}
\end{figure}

\begin{figure}[h]
\begin{center}
\epsfig{file=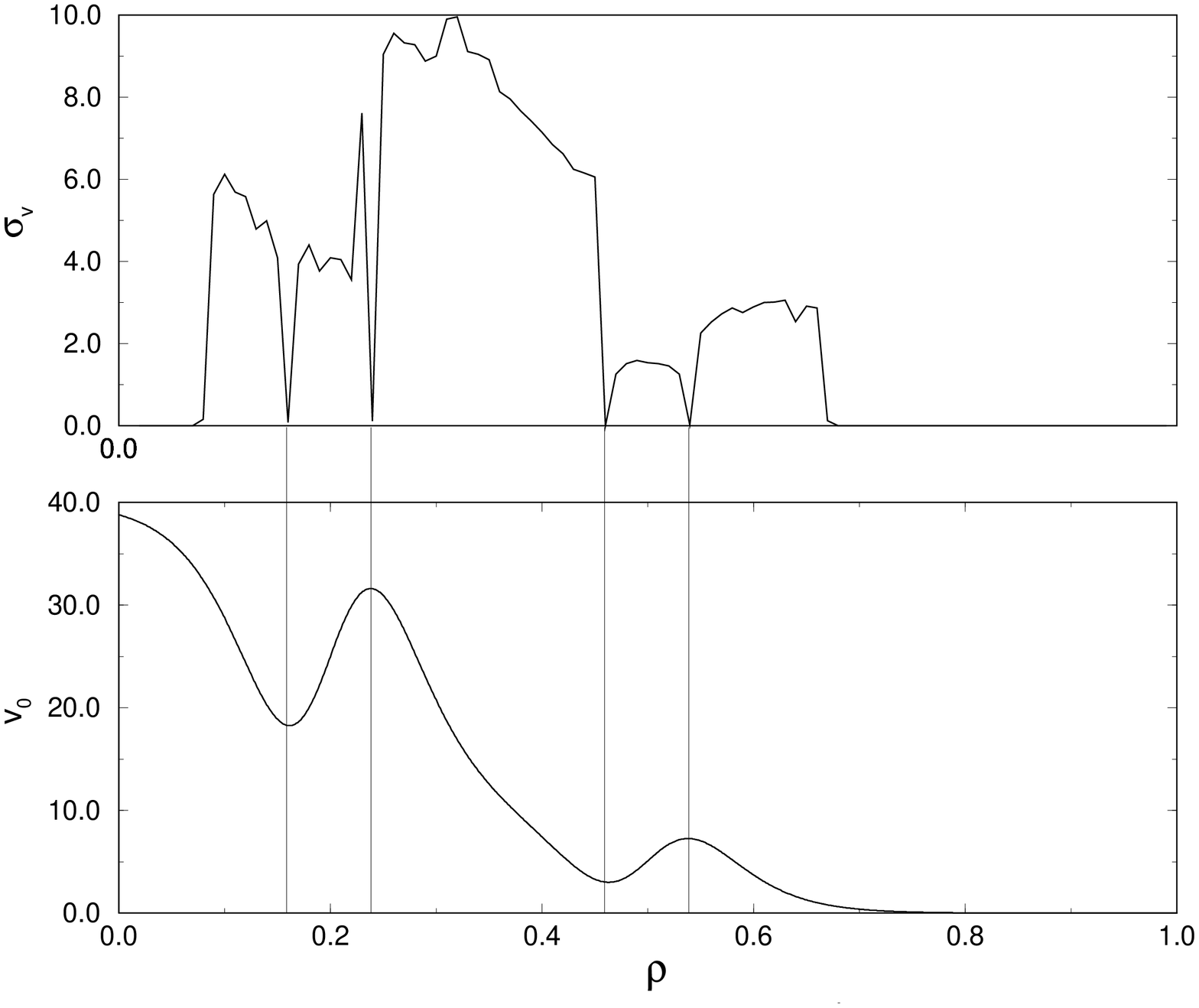,angle=-90,width=7cm}
\epsfig{file=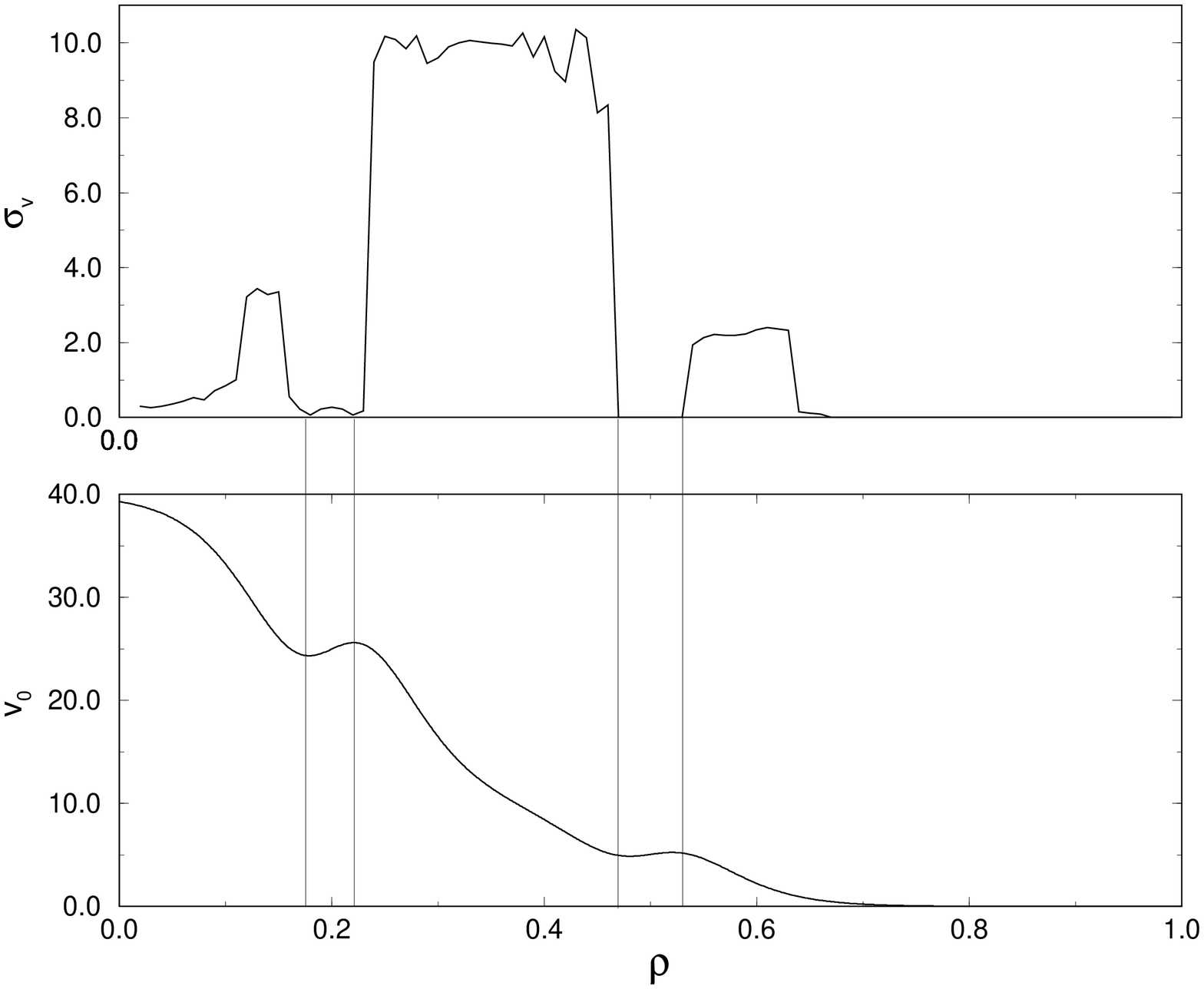,angle=-90,width=7cm}
\caption{\label{velvar_comp} Shown are comparisons of the $\sigma_v$
and $v_0$ as functions of $\rho$ for $\tau= 10$ s (left) and  $\tau= 5$ s
 (right) for the case of the plateau function. Stability is encountered where 
$dv_0/d\rho$ is small, otherwise the flow is unstable.}
\end{center}
\end{figure}
Traffic modelling as well as traffic measurements have a long-standing
tradition (e.g. Greenshields \cite{green35}, Lighthill and Whitham 
\cite{lighthill55}, Richards \cite{richards55}, Gazis et al. \cite{gazis61},
 Treiterer \cite{treiterer74}, to name just a few). 
In recent years physicists working in this area have
tried to interpret and formulate phenomena encountered in traffic flow
in the language of non-linear dynamics (see, for example, the review
of Kerner \cite{TGF99_kerner} and many of the references cited
therein).  The measured real-world data reveal a tremendous amount of
different phenomena many of which are also encountered in other
non-linear systems.\\
To identify properties of our model equations we apply them to a
closed one-lane road loop. The loop has a length of $L= 10$ km and we
prepare initial conditions close to a homogeneous state with density
$\rho_{in}$ (i.e. same density and velocity everywhere). The system is
slightly perturbed by a sinusoidal density perturbation of fixed
maximum amplitude $\delta \rho= 0.01$ and a wavelength equal to the
loop length. The particles are initially distributed equidistantly,
the weights $n_i$ are assigned according to (\ref{dens}) in order to
reproduce the desired density distribution, and the velocities
corresponding to $U_e(\rho)$ are used. All calculations are performed
using 500 particles.\\
It is important to
keep in mind that the results are only partly comparable to real world
data since the latter may reflect the response of the non-linear system to
external perturbations like on-ramps, accidents etc. which are not
included in the model.\\
In the following the model parameter $\tau$, which determines the
time scale on which the flow tries to adapt on $U_e$, is allowed to
vary. This corresponds to a varying acceleration capability of the
flow due to a  changing vehicle composition (percentage trucks
etc.). This parameter $\tau$ which is typically of the order of
seconds controls a wide variety of different dynamical phenomena. A
similar result has been found in \cite{IDM} for a  microscopic
car-following model.

\subsection{Analysis of the Model Equations}

\subsubsection{Fundamental diagram}

The $x$-isocline, i.e.\ the locus of points in the $x-v$-plane for
which $\frac{dx}{dt}=0$, is found from eq. (\ref{eq1}) to be the
abscissa. The velocity isocline, i.e. the $(x,v)$-points where the
acceleration vanishes can be inferred from eq. (\ref{eq2}).
For the homogeneous and stationary solution one finds the
isocline velocity $v_0$ as a function of $\rho$:
\begin{equation}
v_0 (\rho)\equiv U_e(\rho) + g \tau U''_e(\rho). \label{vel_iso}
\end{equation}
Fixed-points of the flow, defined as intersections of the $x$- and
$v$-isoclines, are thus expected only for densities above $\approx
0.6$, where $v_0$ approaches the abscissa, see Fig.
\ref{force_free_FD}. 
The flow of the homogeneous and stationary solution has no fixed point 
in a strict sense since $U''_e$
becomes extremely small at $\rho=1$, but does not vanish exactly.
However, this could easily be changed by choosing another form of
$U_e$.  Fig.~\ref{force_free_FD} shows these ``force free velocities'' $v_0$ 
(in the homogeneous and stationary limit) together with the ``force free
fundamental diagrams'' (for $\tau= 1, 5, 10$ s) for both investigated forms of
$U_e$.  We expect the
fundamental diagrams (FD) found from the numerical analysis of the
full equation set to be centered around these ``force free fundamental
diagrams''.  While for the $U_e$ of the $K^3$-model the FD
always (i.e. for $\tau= 1, 5, 10$ s) exhibits a simple one-hump
behaviour, eq. (\ref{Vdes_eq}) leads to a one, two or three hump
structure of the FD depending on the time constant $\tau$.  That is
why a stronger dependence of qualitative features on this constant may
be expected for the plateau safe velocity.
Note, however, that even with the conventional $U_e$ the ``force free 
velocity''
 $v_0$ exhibits for $\tau=10$ s two additional extrema at intermediate 
densities (up to four with the plateau function). The implications of these 
additional extrema for the stability of the flow are discussed below. \\

\subsubsection{Stability}

 To get a preliminary idea about the stability regimes of the model it is 
appropriate to perform a {\em  linear stability analysis}. By 
inserting (\ref{vel_iso}) into the
equations of motion we obtain equations that formally look like the
equations of the $K^3$-model:
\begin{equation}
\dot{v}_i= \frac{v_0(\rho)-v_i}{\tau}- \frac{c_0^2}{\rho} 
\frac{\partial \rho}{\partial x}
+ \frac{\mu}{\rho} \frac{\partial^2 v}{\partial x^2},
\end{equation}
the role of $U_e$ now being played by $v_0$. Thus with the appropriate
substitution the {\em linear stability criterion} of \cite{kerner93}
can be used:
\begin{equation}
\frac{dv_0}{d \rho} < - \left[1+\left(\frac{2 \pi}{L}\right)^2 
\frac{\mu \tau}{\rho}\right] \frac {c_0}{\rho}. \label{stab_crit}
\end{equation}
Thus we expect the flow to be linearly unstable in density regimes
where the decline of $v_0$ with $\rho$ is steeper than a given
threshold. Specifically, extrema of the $v_0(\rho)$ are (to linear
order) stable and we therefore expect {\em stable density regions embedded 
in unstable regimes}.

\subsection{Simulation Results}
The previous analytical considerations give a rough idea of what to expect, 
for a more complete analysis, however, we have to resort to a numerical 
treatment of the full equation set. In order to be able to distinguish the
effects resulting from the additional term in the equations of motion from
those coming from the form of $U_e$ we treat two cases separately:
in the first case the conventional form of $U_e$ is used and in the second the 
effects due to a plateau in $U_e$ \cite{kerner93} are investigated.

\subsubsection{Conventional Form of $U_e$}

To obtain a {\bf fundamental diagram} (FD) comparable to measurements
we chose a fixed site on our road loop. We determine averages
over one minute in the following way: 
$\rho_{1m}= p^{-1} \sum_{i=1}^{p} \rho_i$ and
$\j_{1m}= p^{-1} \sum_{i=1}^{p} \rho_i v_i$, where $p$ is the number
of particles that have passed the reference point within the last
minute. The thus calculated FD (Fig. \ref{FD_allTau}, left panel) 
is, as expected, close to a superposition of the
``force-free FDs'', for different values of $\tau$ see Fig. \ref{force_free_FD}, 
left panel.  As in real-world
traffic data in the higher density regimes the flow is not an
unambiguous function of $\rho$, but rather covers a surface given by
the range of $\tau$ in the measured data. Note that many data points in
the unstable regime (see below) exhibit substantially higher flows than 
expected from the "force-free FDs" (see Fig. \ref{force_free_FD}).\\
In certain density ranges the model shows {\bf instability} with respect
to jam formation from an initial slight perturbation. In this regime
the initial perturbation of the homogeneous state grows and finally
leads to a breakdown of the flow into a backward moving jam (Kerner
refers to this state, where vehicles come in an extended region to a
stop, as "wide jam" (WJ) contrary to a "narrow jam" (NJ) which
basically consists only of its upstream and downstream fronts and
vehicles do not, on average, come to a stop; \cite{PRL98}). This
phenomenon, widely known as "jam out of nowhere", is reproducible with
several traffic flow models (e.g. \cite{kerner93,nagel92,krauss98}).
An example of a spontaneously forming WJ accompanied by two NJs is given 
in Fig. \ref{widejam}, left panel, for an initial density of 
$\rho_{in}= 0.25$. It is interesting to note that the initial perturbation
remains present in the system for approximately 15 minutes without 
noticeably growing in amplitude before the flow breaks down.
As in reality the inflow front of the WJ is much steeper than the outflow 
front of the jam.  Note the similarity with the 
jam formation process within the K$^3$-model \cite{kerner93}.\\
To give a global idea in which density regimes congestion phenomena
occur we show in Fig.~\ref{velvar}, left panel,  the
velocity variance $\sigma_v= \sqrt{\frac{1}{N}\sum_i (v_i-\bar{v})^2}$
for given initial densities.
The system is allowed to evolve from its initial state until
$\sigma_v$ converges. If $\sigma_v$ has not converged after a very
long time $T_s$ (= 10000 s) it is assumed that no stationary state
($\sigma_v \approx$ const) can be reached and $\sigma_v$ is taken at
$T_s$.  $N$ denotes the particle number and $\bar{v}$ the average
particle velocity in the system. For low values of $\tau$ (1 and 5 s)
the system shows spontaneous jam formation in a coherent density regime from 
$\sim 0.16$ to $\sim 0.5$, comparable to measured data. For $\tau=10$ s a 
stable regime at intermediate densities surrounded by unstable density regimes
is encountered. This region corresponds to the two close extrema seen in Fig.
\ref{force_free_FD}, first panel.\\
Another widespread phenomenon is the formation of several jams
following each other, so-called {\bf stop-and-go-waves}. This phenomenon is
also a solution of our model equations, see Fig.~\ref{stop_and_go}, left panel.
The emerging pattern of very sharply localized perturbations is found
in empirical traffic data as well (see Fig. 14, detector D7 in
\cite{TGF99_kerner}). \\
A very interesting phenomenon happens towards the upper end of the instability
range ($\rho_{in}\sim 0.5$). After the initial perturbation has remained 
present in the system for more than 20 minutes without growing substantially
in amplitude, see Fig.~\ref{plateau1}, suddenly a sharp velocity spike appears
 at $t= 1750$ s that broadens in the further evolution until the system
has separated into two phases: a totally queued phase, where the velocity 
vanishes on a distance of several kilometers, and a homogeneous high velocity 
phase, both separated by a shock-like transition. We refer to these  states 
with homogeneous velocity plateaus separated by shock fronts as 
{\bf Mesa states}.

\subsubsection{$U_e$ with Plateau}

\begin{figure}[h]
\begin{center}
\epsfig{file=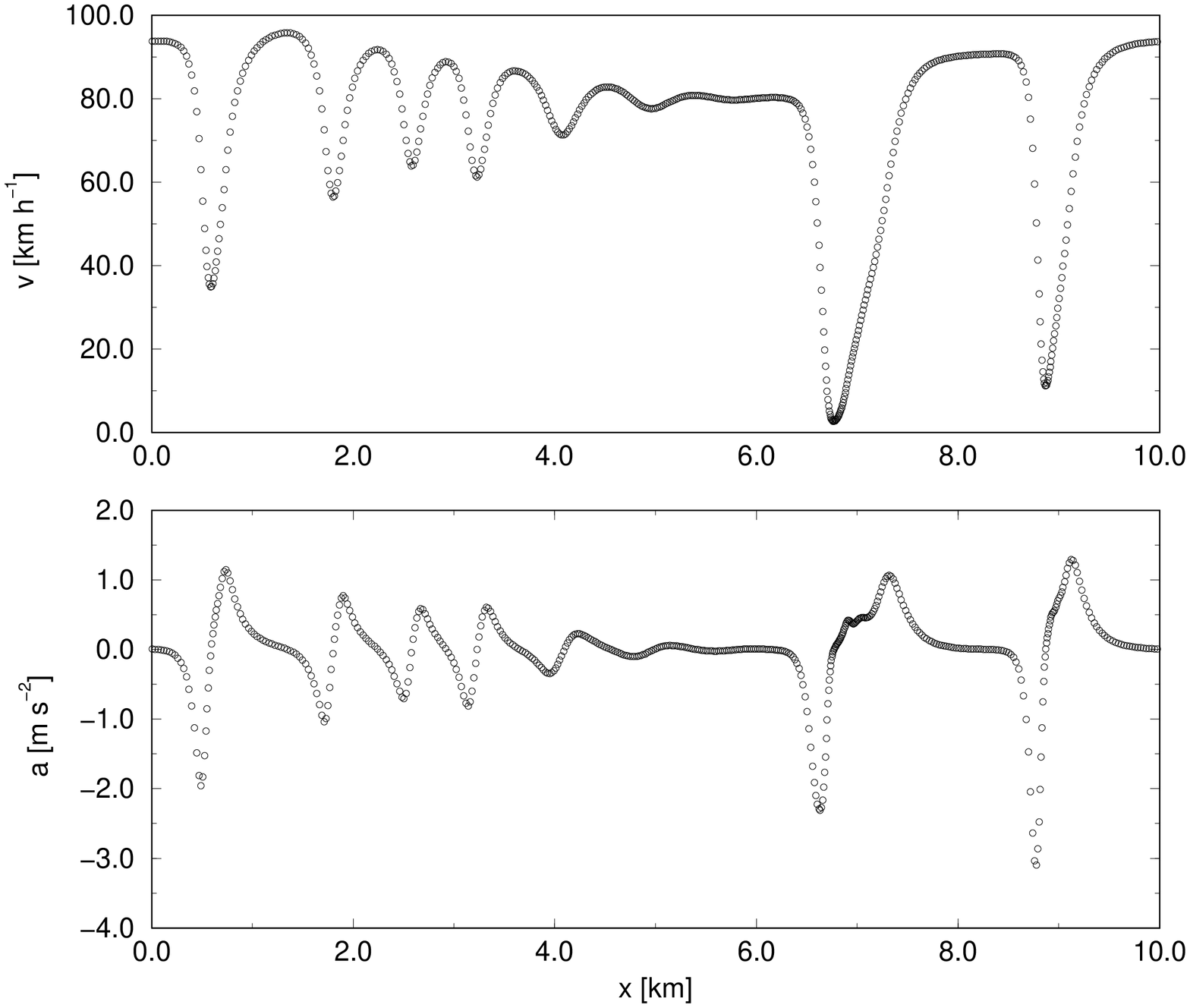,angle=-90,width=8cm}
\caption{\label{acc} Velocities and accelerations in a violently congested
state (480 s after simulation start, $\rho_{in}= 0.25, \mu= 50$
ms$^{-1}, \tau= 3$ s).  The encountered accelerations are always and
everywhere in the range expected from experimental traffic data.}
\end{center}
\end{figure}

The numerically determined {\bf fundamental diagrams} for the case with 
plateau is shown in 
Fig.~\ref{FD_allTau}, right panel. The additional extrema expected
from the "force-free velocity"  $v_0$ are visible in the data points.
We therefore conclude that {\em if a pronounced 
plateau in $U_e$ really does exist, additional extrema should appear in the 
measured fundamental diagrams, at least for flows with poor acceleration 
capabilities, i.e. large $\tau$'s}.\\
Also with the plateau function the system shows {\bf spontaneous jam 
appearance}. The formation of an isolated, stable WJ is displayed in 
Fig.~ref{widejam}, right panel. With a  change in the parameter $\tau$ 
(10 s rather than 5 s as in Fig.~\ref{widejam}) one finds a more 
complicated pattern with one WJ that coexists for a long time with 
constantly emerging and disappearing NJs, see Fig. \ref{jamsync}.\\
The global stability properties for the case with plateau are shown in 
Fig.~\ref{velvar}, right panel. As expected from the linear stability 
analysis (see eq. (\ref{stab_crit}) and Fig.~\ref{force_free_FD}, right panels) we find
alternating regimes of stability and instability rather than one coherent  
density range where the flow is prone to instability.
For low ($\rho \lesssim 0.1$) and very high density
($\rho \gtrsim 0.7$), initial perturbations decrease in amplitude,
i.e.\ the system relaxes towards the homogeneous state. In between
these density perturbations may grow and lead to spontaneous structure
formation of the flow. The stable regions within unstable flow are
found around densities for which $\frac{dv_0}{d\rho}= 0$. This is displayed
for two values of $\tau$ in Fig. \ref{velvar_comp}. \\ 
The {\bf accelerations} in the model were never found to exceed $\sim 4$
ms$^{-2}$ for negative and $\sim 1.5$ ms$^{-2}$ for positive signs and
thus agree with accelerations from real-world traffic data (for both 
forms of $U_e$). For
reasons of illustration Fig. \ref{acc} displays velocities and the
corresponding accelerations at one time slice of a simulation
($\rho_{in}= 0.25, \mu= 50$ ms$^{-1}, \tau= 3$ s) for $U_e$ according to
eq. (\ref{Vdes_eq}).

Also the plateau function allows for {\bf stop-and-go-waves}, see 
Fig.~\ref{stop_and_go}, right panel.
The shown evolution process is close to
what Kerner \cite{PRL98} describes as general features of
stop-and-go-waves: initiated by a local phase transition from free to
synchronized flow, numerous well localized NJs emerge, move through
the flow and begin to grow. One part of the NJs propagates in the
downstream direction (see e.g.\ the perturbations located at $\sim 2$
km at $t= 400$ s) while the rest (at $t=400$ s at $\sim 8$ km) move
upstream. Once the first WJ has formed after approximately 500 s the
NJs start to merge with it. This NJ-WJ merger process continues until
a stationary pattern of three WJs has formed (at around 1000 s; not
shown) which moves with constant velocity in upstream direction. The
distance scale of the downstream fronts of these self-formed WJs is in
excellent agreement with the experimental value of 2.5 - 5 km
\cite{PRL98}.\\
We found for the conventional form of $U_e$ a separation into different
homogeneous velocity phases that we called {\bf Mesa states}. This feature
is also present if the plateau function is used.
In Fig. \ref{plateau1}, right panel, the initial perturbation organizes 
itself into  different platoons of homogeneous velocities. These platoons
are separated by sharp, shock-like transitions and form a stationary
pattern that moves along the loop without changing in shape.\\
\begin{figure}[h]
\begin{center}
\epsfig{file=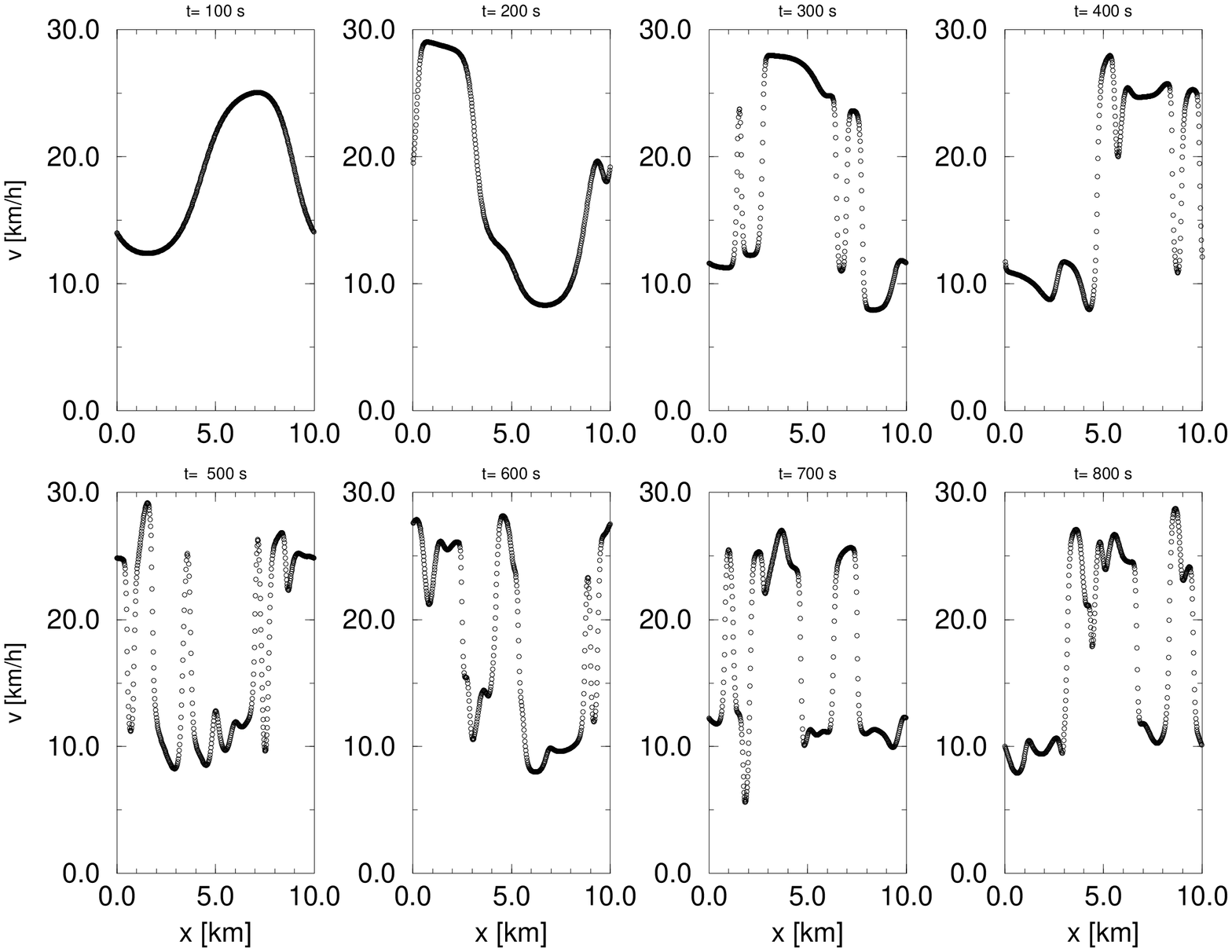,angle=-90,width=10cm}
\caption{\label{plateau2} "Mesa-effect 2": the composition of the flow (e.g.
the fraction of trucks) plays a crucial role for the stability of the 
velocity plateaus ($\rho_{in}= 0.50, \mu= 50$ ms$^{-1}, \tau= 12$ s).
The same initial conditions for the plateau function do not lead to a
behaviour that is qulitatively different from Fig.~\ref{plateau1}, left panel.}
\end{center}
\end{figure}
The relaxation term in eq. (\ref{modeq}) plays a crucial
role for the stabilisation of this pattern. If, for example, the
relaxation time $\tau$ is increased (see Fig. \ref{plateau2}) and thus
the importance of the relaxation term is reduced, the system is not
able to stabilize the velocity plateau. It seems to be aware of these
states, but it is always heavily disturbed and never able to reach a
stationary state. Again, the composition of the traffic flow plays via
$\tau$ {\em the} crucial role for the emerging phenomena.

\clearpage

\section{Summary}

Starting from the assumption that a safe velocity $U_e(\rho)$ exists
towards which drivers want to relax by anticipating the density ahead
of them, we motivate a set of equations for the temporal evolution of
the mean flow velocity. The resulting partial differential equations
possess a Navier-Stokes-like form, they extend the well-known
macroscopic traffic flow equations of K\"uhne, Kerner and Konh\"auser
by an additional term proportional to the second derivative of
$U_e(\rho)$. Motivated by recent empirical results, we explore, in
addition to the new equation set, also the effects of a $U_e$-function
that exhibits a plateau at intermediate densities. The results are compared to
the use of a conventional form of $U_e$.  \\
These fluid-like equations are solved using a Lagrangian particle scheme, that
formulates density in terms of particle properties and
evaluates first order derivatives  analytically by means of cubic spline
interpolation and second order derivatives by equidistant finite 
differencing of the splined
quantities. The continuity equation is fulfilled automatically by
construction. This method is able to 
follow the evolution of the (in some ranges physically unstable)
traffic flow in a numerically stable way and to resolve emerging 
shock-fronts accurately without any spurious oscillations.\\ 
The presented model shows for both investigated forms of $U_e$ a large 
variety of phenomena that are well-known from real-world traffic data.  
For example, traffic flow is
found to be unstable with respect to jam formation initiated by a subtle
perturbation around the homogeneous state.  As in reality,
stable backward-moving {\em wide jams} as well as sharply localized
{\em narrow jams} form. These latter ones move through the flow
without leading to a full breakdown until they merge and form wide
jams. The distance-scale of the downstream fronts of these self-formed
wide jams is in excellent agreement with the empirical values. The
encountered accelerations are in very good agreement with measured
values. For the $U_e$ with plateau we also find states where a stable wide
jam coexists with narrow jams that keep emerging and disappearing without ever
leading to a breakdown of the flow, properties that are usually attributed
to the elusive state of "synchronized flow". 
Another, striking phenomena is encountered that we call the
"Mesa-effect": the flow may organize into a state, where platoons of
high and low velocity follow each other, separated only by a very
sharp, shock-like transition region. This pattern is found to be
stationary, i.e. it moves forward without changing shape. 
One may speculate, that these mesa states are
related to the minimum flow phase found in the work using the ASEP as
a model for traffic flow \cite{popkov99}.\\
In other
regions of parameter space the flow is never able to settle into a
stationary state. Here wide jams and a multitude of emerging moving or
disappearing narrow jams may coexist for a very long time. Again, it may be 
presumeded that in these cases the system displays
deterministic chaos, however, we did not check this beyond any doubt.\\
%
The basic effect of the new interaction term is to make the "force-free
velocity", which essentially determines the shape of the fundamental diagram,
sensitive to the relaxation parameter $\tau$. For large values of $\tau$ 
additional extrema in the "force-free flows" are introduced and a stability
analysis shows that that the flows are stable against perturbations in the
vicinity of these extrema. This leads to the emergence of alternating 
regimes of stability and instability, the details of which depend on the 
shape of $U_e$. We find that if a pronounced plateau in $U_e$ really does 
exist, it should appear in the measured fundamental diagrams, at least 
for flows with poor acceleration capabilities, i.e. large $\tau$'s.\\
The crucial parameter besides density which determines the dynamic
evolution of the flow and all the related phenomena is the relaxation
time $\tau$. Since this parameter governs the time scale on which the
flow tries to adapt to the desired velocity $U_e$, we may interpret it
as a measure for the flow composition (fraction of trucks
etc.).  It is this composition that  determines 
whether/which structure formation takes place, whether the system relaxes
into a homogeneous state, forms isolated wide jams or a multitude
of interacting narrow jams.

To conclude, this work shows that a surprising richness of phenomena
is encountered if one allows for a slight change
of the underlying traffic flow equations. Further work is
needed in order to extend the qualitative description undertaken in
this work and to find more quantitative relationships between the
traffic flow models and reality.


\begin{thebibliography}{99}

 \bibitem{Benz90} W.~Benz, in The numerical Modelling of Nonlinear
  Stellar pulsations, Kluwer Academic Publishers, Dordrecht (1990) 

\bibitem{gazis61} D.~C.~Gazis, R.~Herman, and R.~W.~Rothery, Nonlinear
  Follow--the--Leader Models of Traffic Flow, Oper. Res. {\bf 9} 545
  --567 (1961).
  
\bibitem{green35} B.~D.~Greenshields, Studying Traffic Capacity by New
  Methods, Civil Engineering, 1935.
        
\bibitem{IDM} M.~Treiber, A.~Hennecke, and D.~Helbing, Congested
  traffic states in empirical observations and microscopic
  simulations, Physical review E {\bf 62}, 1805 -- 1824 (2000).
 
\bibitem{kerner93} B.~S.~Kerner and P.~Konh\"auser, Cluster effect in
  initially homogeneously traffic flow, Phys. Rev. E {\bf 48(4)}, 2335
  (1993)
  

\bibitem{krauss98} S.~Krau{\ss}, P.~Wagner and C.~Gawron, Metastable
  states in a microscopic model of traffic flow, Phys. Rev. E {\bf
    55}, 5597--5605 (1997).

\bibitem{kuehne84} R.~D.~K\"uhne, Macroscopic Freeway Model for dense
  traffic, Proceedings of the 9th International Symposium on
  Transportation and Traffic Theory, p. 21 (1984).
  
\bibitem{lighthill55} M.~J.~Lighthill and G.~B.~Whitham, On kinematic
  waves: a Theory of Traffic on long crowded Roads, Proceedings of the
  Royal Society, {\bf A 229}, 317 (1955)
  
\bibitem{nagel92} K.Nagel and M. Schreckenberg, A cellular automaton
  model for freeway traffic, {\em J.Physique I}, 2:2221 (1992)
  
\bibitem{payne71} H.J. Payne, Models fro freeway traffic and control, {\em Math. Models Pub. Sys.}, Simul. Council Proc., {\bf 28}, 51 (1971)
  
\bibitem{popkov99} V.~Popkov and G.~M.~Schuetz, Steady state
  selection in driven diffusive systems with open boundaries,
  Europhys. Lett. {\bf 48}, 257 (1999)
  
\bibitem{PRL98} B.S. Kerner, Experimental features of
  Self-organisation in traffic Flow, Phys. Rev. Lett. {\bf 81}, 3797
  (1998)

\bibitem{richards55} P.~I.~Richards, Shock Waves on the Highway, Oper.
  Res. {\bf 3}, 42--51 (1955).

\bibitem{rosswog99} S. Rosswog, P. Wagner, ``Car-SPH'': A Lagrangian Particle
Scheme for the Solution of the Macroscopic traffic Flow Equations,
  Traffic and Granular Flow 99, p. 401
  
\bibitem{TGF99_kerner} B.S. Kerner, Phase transitions in traffic flow,
  Traffic and Granular Flow 99, p. 253
  
\bibitem{treiterer74} J.~Treiterer, J.~A.~Myers, The Hysteresis
  Phenomenon in Traffic Flow, in D.~J.~Buckley (ed.), Proc. of the
  Sixth Intern. Symp. on Transportation and Traffic Theory, Elsevier
  1974.
  
\end{thebibliography}
\end{document}